\documentclass[aps,pra,superscriptaddress,twocolumn]{revtex4}
\usepackage{graphicx,amsmath,amssymb,amsfonts,latexsym,color,dcolumn,bm}

\newcommand{\beq}{\begin{equation}}
\newcommand{\eeq}{\end{equation}}
\newcommand{\bea}{\begin{eqnarray}}
\newcommand{\eea}{\end{eqnarray}}
\providecommand{\abs}[1]{\left\lvert#1\right\rvert}

\providecommand{\bra}[1]{\langle #1 \rvert}
\providecommand{\ket}[1]{\lvert #1 \rangle}

\usepackage{amsfonts,amssymb}
\usepackage{dsfont}
\usepackage{physics}
\usepackage{tocvsec2}

\usepackage{appendix}

\usepackage{tikz}

\usepackage{amsmath}
\usepackage{bbold}
\usepackage{hyperref}
\hypersetup{
     colorlinks=true,      
    linkcolor=blue,        
    citecolor=blue,        
    filecolor=magenta, 
    urlcolor=black          
}

\usepackage{comment}

\begin{document}

\title{Producing delocalized frequency-time Schr\"odinger cat-like states with HOM interferometry}
\author{N. Fabre\footnote{nicolas.fabre@univ-paris-diderot.fr} }
\affiliation{Laboratoire Mat\'eriaux et Ph\'enom\`enes Quantiques, Sorbonne Paris Cit\'e, Universit\'e de Paris, CNRS UMR 7162, 75013 Paris, France}
\author{J. Belhassen}
\affiliation{Laboratoire Mat\'eriaux et Ph\'enom\`enes Quantiques, Sorbonne Paris Cit\'e, Universit\'e de Paris, CNRS UMR 7162, 75013 Paris, France}
\author{A. Minneci}
\affiliation{Laboratoire Mat\'eriaux et Ph\'enom\`enes Quantiques, Sorbonne Paris Cit\'e, Universit\'e de Paris, CNRS UMR 7162, 75013 Paris, France}
\author{ S. Felicetti}
\affiliation{Istituto di Fotonica e Nanotecnologie, Consiglio Nazionale delle Ricerche (IFN-CNR), Piazza Leonardo da Vinci 32, 20133 Milano, Italy}
 \author{A. Keller}
\affiliation{Laboratoire Mat\'eriaux et Ph\'enom\`enes Quantiques, Sorbonne Paris Cit\'e, Universit\'e de Paris, CNRS UMR 7162, 75013 Paris, France}
 \author{M.I. Amanti}
\affiliation{Laboratoire Mat\'eriaux et Ph\'enom\`enes Quantiques, Sorbonne Paris Cit\'e, Universit\'e de Paris, CNRS UMR 7162, 75013 Paris, France}
 \author{F. Baboux}
\affiliation{Laboratoire Mat\'eriaux et Ph\'enom\`enes Quantiques, Sorbonne Paris Cit\'e, Universit\'e de Paris, CNRS UMR 7162, 75013 Paris, France}
 \author{T. Coudreau}
\affiliation{Laboratoire Mat\'eriaux et Ph\'enom\`enes Quantiques, Sorbonne Paris Cit\'e, Universit\'e de Paris, CNRS UMR 7162, 75013 Paris, France}
 \author{S. Ducci}
\affiliation{Laboratoire Mat\'eriaux et Ph\'enom\`enes Quantiques, Sorbonne Paris Cit\'e, Universit\'e de Paris, CNRS UMR 7162, 75013 Paris, France}
 \author{P. Milman}
\affiliation{Laboratoire Mat\'eriaux et Ph\'enom\`enes Quantiques, Sorbonne Paris Cit\'e, Universit\'e de Paris, CNRS UMR 7162, 75013 Paris, France}

\date{\today}
\begin{abstract}
In the late 80's, Ou and Mandel experimentally observed signal beatings by performing a non-time resolved coincidence detection of two photons having interfered in a balanced beam splitter [Phys. Rev. Lett {\bf 61}, 54 (1988)]. In this work, we provide a new interpretation of the fringe pattern observed in this experiment as the direct measurement of the chronocyclic Wigner distribution of a frequency Schr\"odinger cat-like state produced by local spectral filtering. Based on this analysis, we also study time-resolved HOM experiment to measure such frequency state.
\end{abstract}
\vskip2pc

\maketitle

\section{Introduction}

\begin{figure}
\includegraphics[width=0.45\textwidth]{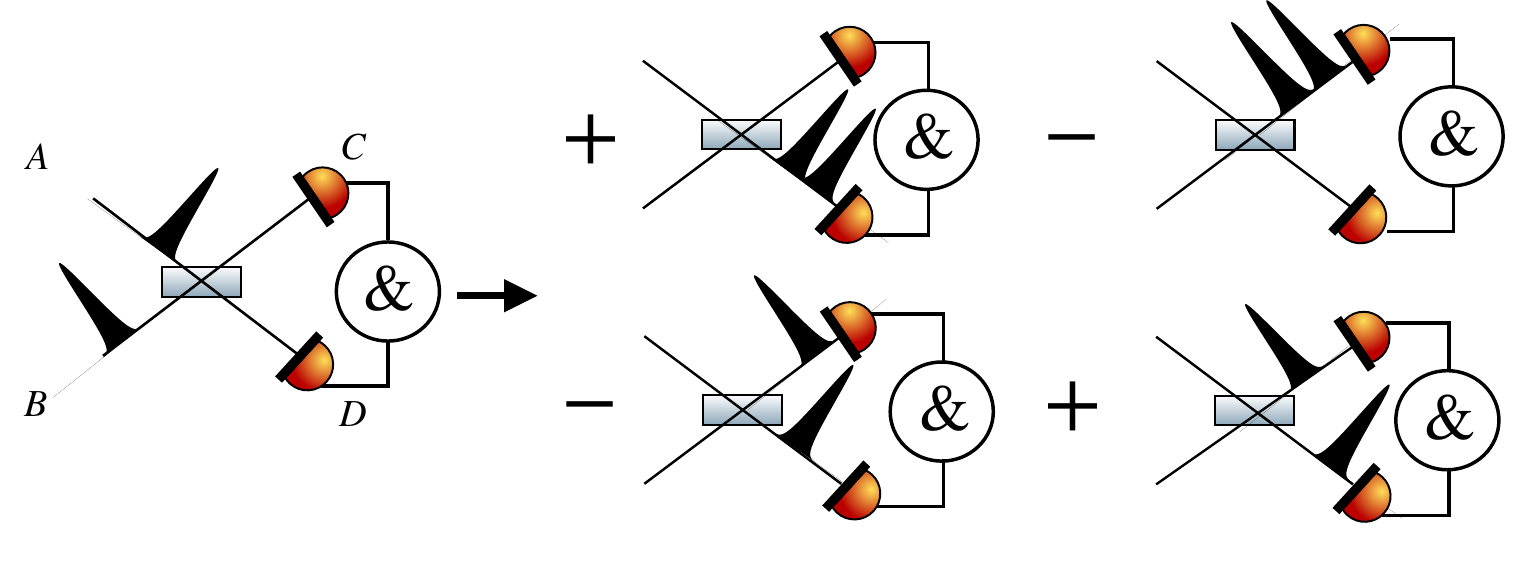}
\caption{\label{FigureHom} Two indistinguishable photons are sent into the paths A and B, one of the photons is delayed of a time $\tau$. After the beam-splitter, four possible cases are possible. Then coincidence measurement is performed, the only events recorded are those where there is one single photon at each spatial port C and D.}
\end{figure}

The Hong-Ou-Mandel experiment \cite{HOM} is one of the foundational experiments of quantum optics illustrating, for the first time, the symmetry properties of particles wave functions. In this experiment, two photons produced by spontaneous parametric down conversion (SPDC) are sent to different paths, one of which has a dephasing element with respect to the other. We will consider here that this optical path difference depends on time delay $\tau$ between the two arms. In the original version of the experiment \cite{HOM}, two indistinguishable photons are combined in a beam-splitter, after which coincidence measurements are performed. Since photons are bosons, one expects that, if they are identical, they should systematically bunch, meaning that both will take the same output, which can be either the C or the D path in Fig.~\ref{FigureHom}. 
Bunching will happen with equal probability in each arm, if the beam-splitter is balanced, which is the case  considered here. A consequence of bunching is that it eliminates coincidence detection. The coincidence counting as a function of the time delay $\tau$ will exhibit a characteristic continuous dip when the photons are indistinguishable at $\tau=0$, that smoothly disappears when $\tau$ moves away from zero.  However, photons distinguishability depends on several degree of freedom, as polarization, temporal mode \cite{TemporalHOM}, spectral mode \cite{SpectralHOM}  or arrival time \cite{Mosley}.

Using the manipulation of photonic degrees of freedom, such as the transverse spatial modes and polarization \cite{SteveHOM}, one can also use the HOM interferometer to demonstrate anti-bunching, revealed by and anti-dip, that can be associated to a fermionic-like behaviour \cite{integrated2,fermionic2} and is also an entanglement witness \cite{Eckstein08,Tom,integrated1,Fedrizzi2009}.  

Ou and Mandel in \cite{OM} proposed a different version of the HOM experiment, adding frequency filters just before each detector.  The observed coincidence rate as a function of the delay $\tau$  presents an oscillation pattern. Remarkably, the beating period in such two-photon interference experiment is shorter than the time resolution of the detectors, which is a consequence of the fact that the described experiment evidences path interference rather than time interference.


In the present contribution we will present a new interpretation of Ou-Mandel (OM)-type experiments in terms of frequency-time phase distribution leading to a new way of engineering and detecting frequency entangled states by spectral post-selection. Our work provides an interpretation to the number of oscillations appearing in the OM experiment and and it opens the way to the application of the engineered states in different quantum protocols. 
Our work builds on the results obtained in \cite{Tom}, where it has been shown that the HOM experiment is the direct measurement of the Chronocylic Wigner distribution associated to a collective variable of the photon pair.  Following these lines, we develop a general interpretation of OM-type experiments using filters of arbitrary width in frequency, allowing to elucidate the fringe pattern appearing in \cite{OM,discretecolor, catsilicon}, as well as the interference patterns reported in \cite{Rempe}.  More specifically, we will show that the oscillations pattern in the measured coincidence probability thanks to a HOM interferometer corresponds to the interference pattern appearing in the chronocylic Wigner distribution of a two-photon Schr\"odinger cat (SC)-like state or called frequency-time cat-like state. Either the frequency-time cat-like state is produced by spectral post-selection in a non-resolved time OM experiment  as in \cite{OM} or the signature of such state is detected with a time-resolved HOM experiment \cite{Rempe}.

In order to illustrate these concepts and provide results of numerical simulation, we present the case study of a transversally pumped semiconductor AlGaAs waveguide generating counterpropagating signal and idler photons \cite{Orieux11,Orieux2}.

This paper is organized as follows. We start in Sec.~\ref{section2} by defining the frequency-time cat-like state in Sec.~\ref{definitioncat} and the reinterpretation of the OM experiment in terms of that state in Sec.~\ref{reinterpretation}. We propose various ways to modify such measurement apparatus so that the full chronocyclic Wigner distribution structure of a frequency-time cat state can be revealed in Sec.~\ref{effectfilter} and Sec.~\ref{quantumeraser}.
In Sec.~\ref{effecttime}, we study the time-resolved HOM experiment reported in \cite{Rempe}, and reinterpret the spatial beating obtained  by measuring the joint temporal distribution as the production of a frequency-time cat-like state. Nevertheless, the detection is by nature very different and corresponds in this experiment to the measurement of the marginal of the chronocyclic Wigner distribution whereas it corresponds to its cut at the zero frequency in the non-resolved time detection of the HOM experiment. This last experiment can be related to a biphoton Young experiment, which is fully investigated in Sec.~\ref{sectionthreeB}.
Finally in Sec.~\ref{sectionfour}, we propose an experimental scheme to perform quantum gates in the frequency-time variables so as to measure different marginals of the frequency-time distribution of a frequency-time cat-like state.

\section{Production of frequency-time cat-like state by post-selection }\label{section2}

\subsection{Definition of frequency-time cat-like} \label{definitioncat}
We start by considering a time-frequency coherent state, introduced in \cite{fabregkp} and is defined by the wavefunction:
\begin{equation}\label{coherenttime}
\ket{\omega_{1},\tau}=\int \text{d}\omega f_{\omega_{1}}(\omega)e^{i\omega\tau}\ket{\omega},
\end{equation}
where $f_{\omega_{1}}(\omega)$ is a Gaussian function centered at $\omega_{1}$ of width $\Delta\omega$. $\ket{\omega}$ denotes a single photon state at frequency $\omega$. We define a frequency-time cat-like state as a single photon state having a frequency distribution composed of two peaks centered around the frequencies $\omega_{a}+\omega_{1}$ and $\omega_{a}-\omega_{1}$. The corresponding wave function can be expressed as:
\begin{multline}\label{frequencytimeSC}
\ket{\psi}=\frac{1}{\sqrt{2}}\int \text{d}\omega (f_{\omega_{a}+\omega_{1}}(\omega)e^{i\omega\tau}+ f_{\omega_{a}-\omega_{1}}(\omega)e^{-i\omega\tau})\ket{\omega}\\
=\frac{1}{\sqrt{2}}(\ket{\omega_{a}+\omega_{1},\tau}+\ket{\omega_{a}-\omega_{1},-\tau}).
\end{multline}
For simplicity, we set $\omega_{a}=0$. This definition is mathematically analogous to the usual Schr\"odinger cat state in position-momentum phase space $(x,p)$, defined as the linear superposition of two macroscopically distinguishable coherent states, $\ket{\psi}=\frac{1}{\sqrt{2}}(\ket{\alpha}+\ket{-\alpha})$. This can be simply seen by noticing that the coherent state $\ket{\alpha}$ can be expressed in the position basis: $\ket{\alpha}=\ket{x_{0},p_0}=\int \text{d}x e^{-(x-x_0)^{2}}e^{ixp_0}\ket{x}$, where $(x_0,p_0)$ is analogous to $(\omega_{1},\tau)$ and is hence mathematically analogous to Eq. (\ref{coherenttime}).  To sum up, standard Schr\"odinger cats are single-mode multi-photon states composed of the superposition of two coherent states. On the other hand, time-frequency cat-like states are given by a single-photon distributed on a continuum of frequency modes, whose frequency distribution is the superposition of two Gaussian peaks.  In addition, we can see that both the coherent state $\ket{x_{0},p_{0}}$ and the state $\ket{\omega_{1},\tau}$ form an over-complete basis and obeys the relation $\bra{\omega_{1},\tau_{1}}\ket{\omega_{2},\tau_{2}}=e^{-(\tau_{1}-\tau_{2})^{2}/4}e^{-(\omega_{1}-\omega_{2})^{2}/4}e^{i(\tau_{1}-\tau_{2})(\omega_{1}+\omega_{2})/2}$. The values $(x_{0},p_{0})$ and $(\omega_{1},\tau)$ correspond to the center of the Gaussian distribution in phase space and the chronocyclic one respectively. Some differences between these two states can be noticed. The free evolution trajectory in the quadrature position-momentum phase space of a coherent state $\ket{\alpha}$ is a circle and the average number of photons is given by $\abs{\alpha}^{2}$. On the other hand, the average value of the photon number of the time-frequency coherent state is one, since the state is a single photon. The trajectory of its free evolution in time-frequency phase space is a translation along the time axis. The denomination "coherent" in that case is not related to the photon number statistics of the state, which is sub-Poissonian for a single photon state, but only due to its mathematical analog structure.\\

 In order to complete the mathematical analogy between the frequency basis and the quadrature position one, we can also notice that the frequency-time phase space of a single photon is non-commutative similarly to the quadrature position-momentum phase space as shown in \cite{fabregkp}. As it happens, one can define non-commuting frequency-time displacement operators \cite{fabregkp} which obeys the Weyl algebra \cite{weyl} and hence have a complete mathematical correspondence with the position-momentum quadrature displacement operators. In this way, the need for a time operator is bypassed, even if a "time of arrival" operator has been defined recently \cite{Maccone}.  As a consequence, the frequency-time phase space of a single photon exhibits a paving structure due to the non-commutativity of these displacement operators as is the case for the $(x,p)$ phase space. Such structure disappears in the classical limit $\hbar\rightarrow 0$. It is worth mentioning that the proposed distribution should not be mingled  to the Wigner-Ville distribution \cite{WignerVille} which is currently used in frequency-time classical signal processing. The frequency-time phase space  for high intensity laser is commutative and completely classical. In both representations, either the phase space $(x,p)$ one or the $(\omega,t)$ one, the Wigner distribution associated to a cat state has a characteristic shape consisting of two separated peaks representing one of the two possible classically distinguishable states and an interference pattern along the symmetry axis, as shown in Fig.~\ref{fig: chathom1} (c).\\

\subsection{Reinterpretation of the OM experiment}\label{reinterpretation}

We now provide an original interpretation of the Ou-Mandel experiment \cite{OM} and show how it can be used to produce interesting non-classical frequency states of photon pairs by post-selection.

\begin{figure*}
\includegraphics[width=0.9\textwidth]{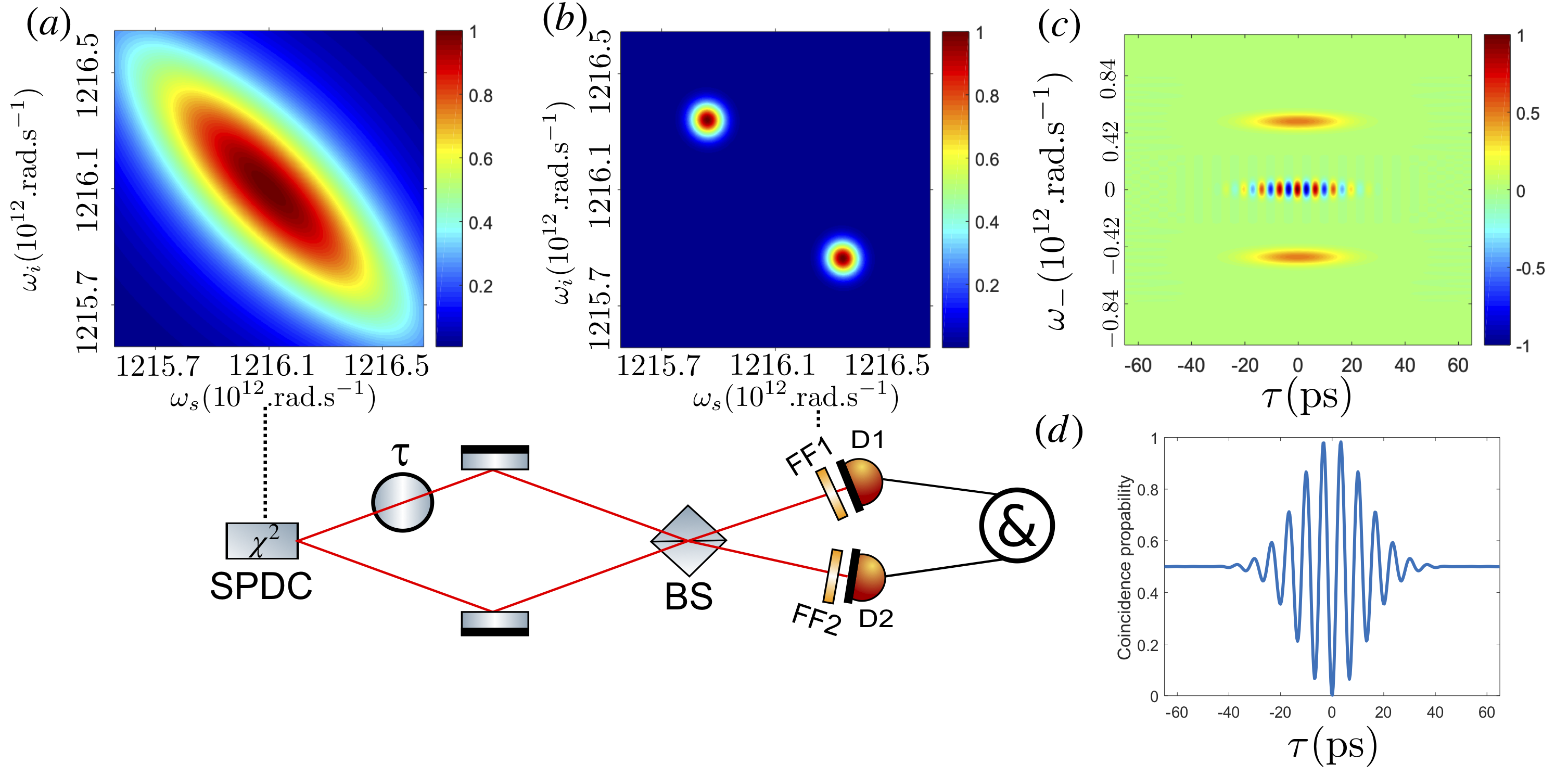}
\caption{\label{fig: chathom1} Sketch of the experimental set-up of the Ou Mandel experiment:  Photon pairs are generated via SPDC in a second order nonlinear crystal. One of the photon undergoes a temporal delay $\tau$, the path of the two photons recombine in a balanced beam-splitter. Spectral filters are placed before the single photon detectors at the output ports of the beam-splitter. The numerical simulations correspond to the state generated in a transversally pumped AlGaAs waveguide: (a) Joint Spectral Amplitude of the frequency anti-correlated state generated by the AlGaAs chip; (b) Joint Spectral Intensity obtained after the action of the beam-splitter and the spectral filters (FF1, FF2) having a width of 50 pm and  a separation of 0.6 nm; (c) corresponding chronocyclic Wigner distribution; (d) coincidence probability corresponding to the cut $\omega_{-}=0$ in (c). }
\end{figure*}

The OM experiment can be summarized as follows: an initial two-photon state is created by  SPDC, whose  Joint spectrum amplitude (JSA), which corresponds to the amplitude of probability of measuring one of the photon pair at frequency $\omega_{s}$ (called the signal) and the other one at frequency $\omega_{i}$ (called the idler) is shown on the left of Fig.~\ref{fig: chathom1} and corresponds to a Gaussian frequency distribution. The wave function can be mathematically expressed as:

\begin{equation}\label{JSA}
\ket{\psi}=\iint \text{d}\omega_{s} \text{d}\omega_{i} \text{JSA}(\omega_{s},\omega_{i})\ket{\omega_{s}} \ket{\omega_{i}},
\end{equation}
where we considered the pump beam to be in the narrow-bandwidth limit  \cite{toolbox}. The  JSA  can be factorized as follows: JSA($\omega_{s},\omega_{i})=f_{+}(\omega_{+})f_{-}(\omega_{-})$ with $\omega_{\pm}=\omega_{s}\pm\omega_{i}$, which is valid for most SPDC sources. The function $f_+$ reflects the energy conservation, and consists of a distribution of width $\xi$, the same one as the pump's, centered at the degeneracy frequency $\omega_{\text{deg}}$. While $f_-$ is related to the phase matching condition and its width is noted as $\Delta$.
We make the approximation $\xi \ll \Delta$ in order to obtain analytical results and impose strict conservation of energy $f_{+}(\omega_{+})=\delta(\omega_{+}-\omega_{\text{deg}})$. For convenience, we will set the frequency $\omega_{\text{deg}}$ to zero.

After their generation, the two photons of the pair are separated into two different paths and are then recombined in a balanced beam splitter as depicted in Fig.~\ref{fig: chathom1}. After the action of the balanced beam-splitter and the post-selecting done by the coincidence measurement, the two-photon state can be expressed as:

\begin{multline}\label{generalstateafter}
\ket{\psi_{\tau}}=\frac{1}{2}[\iint \text{d}\omega_{s} \text{d}\omega_{i}(\text{JSA}(\omega_{s},\omega_{i})e^{-i\omega_{i}\tau}\\-\text{JSA}(\omega_{i},\omega_{s})e^{-i\omega_{s}\tau})\ket{\omega_{s}} \ket{\omega_{i}}],
\end{multline}
as in the usual HOM experiment. We now consider the effect of placing frequency filters before each detector. Each filter can be associated to a projector operator of the type: 
\begin{equation}
\hat F(\omega_{1},\sigma)= \int \text{d}\omega \overline{f_{\omega_{1}}}(\omega)\ket{\omega}\bra{\omega},
\end{equation}
 where $\overline{f_{\omega_{1}}}(\omega)$ is a Gaussian function centered at $\omega_{1}$ of width $\sigma$. In the following, we will consider that the filters have the same spectral width. For standard Gaussian filters,
\begin{equation}\label{filter}
\overline{f_{\omega_{j}}}(\omega_{\alpha})=\frac{1}{\sqrt{2\pi \sigma^{2}}}\text{exp}(-\frac{(\omega_{\alpha}-\omega_{j})^{2}}{2\sigma^{2}}) ,
\end{equation}
with $j=1,2$ and $\alpha=s,i$. The frequency state after spectral post-selection filtering and coincidence detection is:
 
\begin{multline}\label{afterF}
\ket{\psi_{\tau}}=\frac{1}{2}\iint \text{d}\omega_{s} \text{d}\omega_{i}[\text{JSA}(\omega_{s},\omega_{i})e^{-i\omega_{i}\tau}\\-\text{JSA}(\omega_{i},\omega_{s})e^{-i\omega_{s}\tau}]\overline{f_{\omega_{1}}}(\omega_{s})\overline{f_{\omega_{2}}}(\omega_{i})\ket{\omega_{s}} \ket{\omega_{i}}.\end{multline}
The first term of Eq.~(\ref{afterF}) represents the situation where both photons are reflected by the beam-splitter and the second term represents the situation where both are transmitted. We have that the photon pairs passing through the filters are described by a quantum superposition of states having exchanged frequencies. In the case in which both photons are reflected, the signal photon with frequency $\omega_s$ is filtered by the filter centered at $\omega_1$ and the idler with frequency $\omega_i$ is filtered by the filter centered at $\omega_2$. If both are transmitted the opposite occurs. Thus, for $\omega_1-\omega_2 \lesssim \Delta $ the state described by Eq.~(\ref{afterF}) corresponds to what we called a frequency-time cat-like state. These states correspond to two distinguishable states of signal and idler, since $\sigma \ll \Delta$; the quantum superposition comes from the paths interference produced by the beam-splitter.

For easier calculation, we write the product of the frequency function filter in the frequency collective variable:
{\small{\begin{multline}
\overline{f_{\omega_{1}}}(\omega_{s})\overline{f_{\omega_{2}}}(\omega_{i})=\frac{1}{2\pi\sigma^{2}}e^{-\frac{(\omega_{+}-(\omega_{1}+\omega_{2}))^{2}}{4\sigma^{2}}} e^{-\frac{(\omega_{-}-(\omega_{1}-\omega_{2}))^{2}}{4\sigma^{2}}}\\=f_{\omega_{1}-\omega_{2}}(\omega_{-})f_{\omega_{1}+\omega_{2}}(\omega_{+}).
 \end{multline}}}
$f$ is a Gaussian function as $\overline{f}$ (see Eq.~(\ref{filter})) but differ from a factor two in the frequency width.\\
Finally, the wave function after the spectral filtering can be written as:

\begin{multline}\label{postselectfrequencytimeSC}
\ket{\psi_{\tau}}=\frac{A}{2}  \int \text{d}\omega_{-}(f_{-}(\omega_{-})e^{-i\omega_{-}\tau}-f_{-}(-\omega_{-})e^{i\omega_{-}\tau})\\\cross f_{\omega_{1}-\omega_{2}}(\omega_{-}) \ket{\omega_{-},-\omega_{-}},
\end{multline}
with $A=f_{\omega_{1}+\omega_{2}}(0)$. We can notice that the state described by Eq.~(\ref{frequencytimeSC}) and Eq.~(\ref{postselectfrequencytimeSC}) are slightly different, if we consider their chronocyclic Wigner distribution. Indeed the state given by Eq.~(\ref{frequencytimeSC}) has the characteristic shape of the cat state in the frequency-time phase space (see Eq.~(\ref{Wignercat}) in the appendix) but the post-selected state Eq.~(\ref{postselectfrequencytimeSC}) corresponds to a chronocyclic Wigner distribution  with only the interference pattern (see Eq.~(\ref{Wignercatbeat}) in the Appendix).

We can now compute the coincidence probability as a function of the temporal delay:
\begin{equation}\label{coincidencegeneral}
 I(\tau)=\iint \text{d}\omega_{s}\text{d}\omega_{i} \abs{\bra{\omega_{s},\omega_{i}}\ket{\psi_{\tau}}}^{2},
 \end{equation}
where we assumed that the width of the temporal photon wavepacket are very small compared to the temporal resolution of the photodetectors (see the Appendix Sec.~\ref{timeresolution} and Sec.~\ref{effecttime}), which is typically the case for photons produced by SPDC. After integration on the variable $\omega_{+}$, and taking into account that the frequency filters width is much narrower than the phase matching width, $\sigma \ll \Delta$, we obtain:

\begin{multline}\label{Postselectcat}
 I(\tau)=\frac{1}{2}[1-\frac{1}{N}\text{Re}(\int \text{d}\omega_{-} e^{-2i\omega_{-}\tau}\abs{f_{\omega_{1}-\omega_{2}}(\omega_{-})}^{2})],
 \end{multline}
where $N$ is a normalization constant. A last integration gives:

\begin{equation}\label{coincidencecat} 
I(\tau)=\frac{1}{2}[1-e^{-\tau^{2}\sigma^{2}}\text{cos}(2\tau(\omega_{2}-\omega_{1}))].
\end{equation}
Eq.~(\ref{coincidencecat}) shows that the coincidence measurement displays a beating, with a period $\frac{\pi}{\omega_{2}-\omega_{1}}$. For some values of $\tau$, since we observe a coincidence probability that is greater than one half, this beating is a signature of entanglement in frequency variable \cite{Tom, Eckstein08} but is completely independent of the frequency entanglement of the initial state generated by SPDC \cite{OM}. In other words, the Ou and Mandel experiment can be seen as a measurement in a frequency entangled basis and as such, it post-selects an entangled state. In the literature, the term of spatial beating is used to designate such oscillation in the coincidence measurement \cite{OM}.

The above discussion provides a new interpretation of the oscillations shown in Fig.~\ref{fig: chathom1}. Indeed, according to \cite{Tom}, the coincidence probability Eq.~(\ref{Postselectcat}) is the cut of the chronocyclic Wigner distribution $W_{\text{cat}}(\omega_{-},\tau)$ at the frequency $\omega_{-}=0$ (see the appendix Eq.~(\ref{Wignercatbeat})). The measurement procedure corresponds to a measurement in a Schr\"odinger cat basis of the frequency state that has been initially generated by the SPDC crystal Eq.~(\ref{JSA}). We can conclude that the signal obtained by coincidence detection corresponds to the interference term of the frequency-time cat-like state: 
\begin{equation}
I(\tau)=\frac{1}{2}(1-W_{\text{beating}}(0,\tau)).
\end{equation}
Where $W_{\text{beating}}(\omega,\tau)$ corresponds to the chronocyclic Wigner distribution of the phase matching function and is defined in Appendix \ref{Wignercatt}. We verified our results by performing numerical simulations (see Fig.~\ref{fig: chathom1}) on the state generated by SPDC using the experimental parameters for a transversally pumped semiconductor waveguide studied in \cite{Orieux2}. In this device, a pump beam around 775 nm impinging on top of a multilayer AlGaAs waveguide with an incidence angle $\theta$ generates by SPDC two orthogonally polarized signal/idler guided modes around 1.55 $\mu$m. Two Bragg mirrors provide a vertical microcavity for the pump beam increasing the conversion efficiency of the device \cite{Orieux11}. This geometry presents a particularly high degree of versatility in the control of the biphoton frequency correlations via the spatial engineering of the pump beam \cite{integrated2,integrated3}. 
In the numerical simulations reported in Fig.~\ref{fig: chathom1}, the pump beam has a Gaussian intensity profile with a waist 0.2 mm, pulses having a duration of 5 ps. This leads to the generation of a biphoton state with a JSA represented in Fig.~\ref{fig: chathom1} (a). 
The spectral width of the JSA along the $\omega_{+}=\omega_{s}+\omega_{i}$ axis is 1 nm (1.88$\cdot10^{12} \text{rad}.s^{-1}$) and along the $\omega_{-}=\omega_{s}-\omega_{i}$ axis is 0.2 nm (9.42$\cdot10^{12} \text{rad}.s^{-1}$).

In this section, we demonstrate that even starting from frequency anti-correlated photon pairs, which is the most common type of two photon state produced by SPDC, the action of spectral filters allows to post-select Schrödinger cat-like states. The numerical simulation of the chronocyclic Wigner distribution of the filtered JSA and the coincidence probability of the OM experiment are shown in Fig.~\ref{fig: chathom1} (c) and (d) respectively. They both match taking the cut $\omega_{-}=0$ for the chronocyclic Wigner distribution. Notice that the observed oscillations appear irrespectively of the initial state produced by SPDC, since we are dealing here with a post-selection process. Finally, the larger the ratio $(\omega_1 - \omega_2)/\sigma$, the  higher is the number of oscillations in the interference pattern of the frequency-time cat-like state, since this term is a measure of the size of the cat.

\subsection{Effect of the filters parameters on the frequency-time cat-like state}\label{effectfilter}

In this section, we explain how to produce different types of frequency-time cat-like states by spectral post-selection. For such, we will use accordable frequency filters.

 \begin{figure*}
\includegraphics[width=1.0\textwidth]{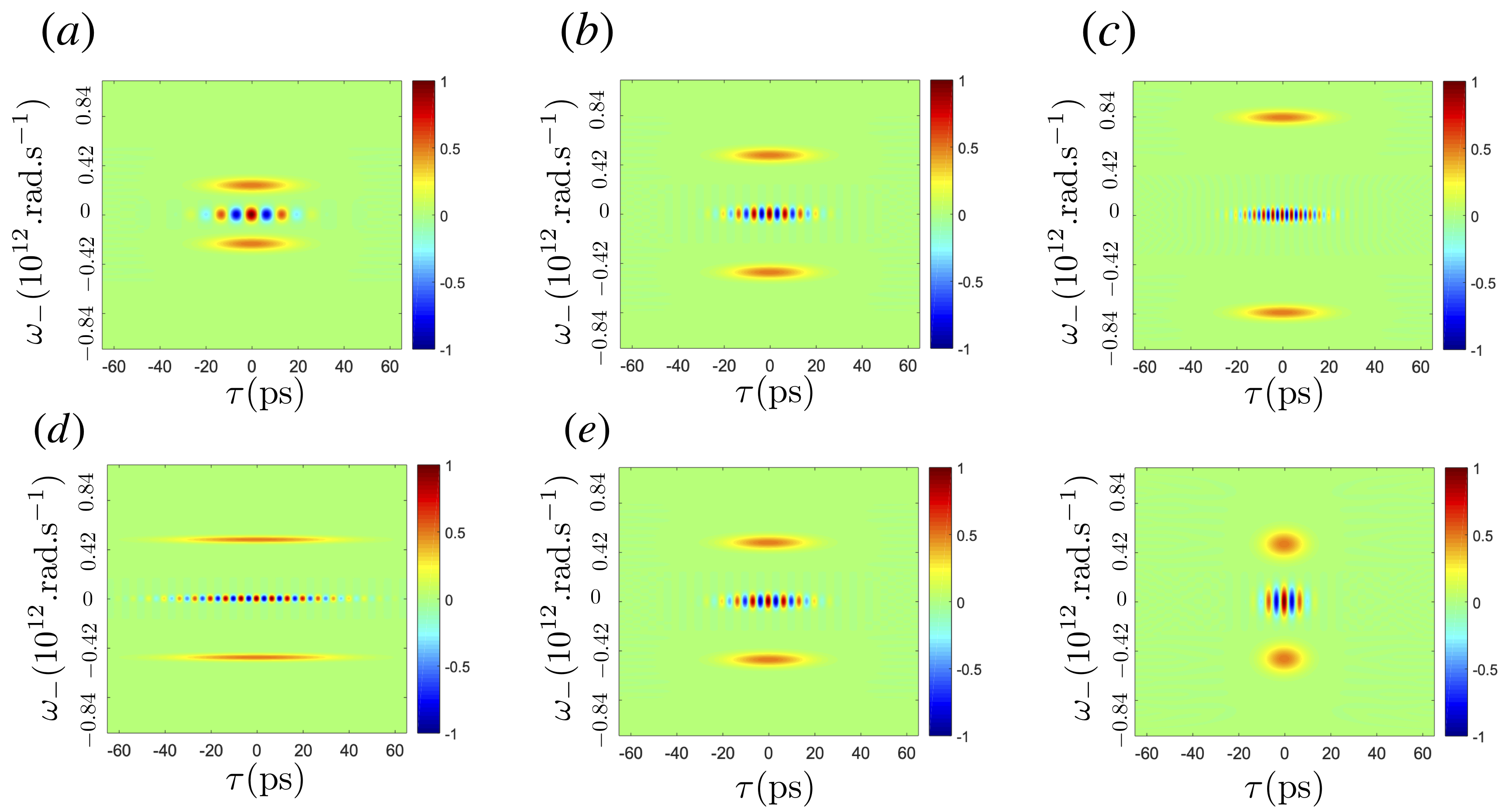}
\caption{\label{filtre1} Chronocylic Wigner distribution of the frequency-time cat-like state. Top: effect of the variation of the frequency separation $\omega_{1}-\omega_{2}$ of the filters of 0.3 nm (6.28$\cdot10^{12} \text{rad}.s^{-1}$) (a), 0.6 nm (3.14$\cdot10^{12} \text{rad}.s^{-1}$) (b) and 1 nm (1.88$\cdot10^{12} \text{rad}.s^{-1}$) (c), for a fixed spectral width of the filters of 50 pm (37.6$\cdot10^{12} \text{rad}.s^{-1}$). Bottom: effect of the variation of the spectral width of the filters (25 pm (75$\cdot10^{12} \text{rad}.s^{-1}$) (d), 50 pm (37.6$\cdot10^{12} \text{rad}.s^{-1}$) (e) and 100 pm (18$\cdot10^{12} \text{rad}.s^{-1}$) (f)) for a fixed spectral separation of 0.6 nm between the central wavelengths of the filters.}
\end{figure*}

We start by studying the effect of the variation of the central frequency $\omega_{1}$ and $\omega_{2}$ of the filters while their width is fixed to 50 pm (37.6$\cdot10^{12} \text{rad}.s^{-1}$), see Fig.~\ref{filtre1} (a),(b),(c). If the central frequency of the filters are equal, we do not obtain an interference pattern along the $\tau$ axis but rather a Gaussian function which is JSA. In that case, the photons are spectrally indistinguishable. When the central frequencies of the filters increases, the period of the beating oscillation $T=\frac{\pi}{\omega_{1}-\omega_{2}}$  (see Eq.~(\ref{coincidencecat}))  decreases and the number of oscillations increases.\\

We then investigate the influence of the spectral width of the filters, keeping fixed the frequency separation $\omega_{1}-\omega_{2}$ of the filters at $0.6$ nm (3.14$\cdot10^{12} \text{rad}.s^{-1}$), see Fig.~\ref{filtre1} (d),(e), (f). As the width decreases from 100 pm (18$\cdot10^{12} \text{rad}.s^{-1}$) to 25 pm (75$\cdot10^{12} \text{rad}.s^{-1}$), the coherence time of the wave packet increases, as a consequence of the Fourier-transform relation between frequency and time.

\subsection{Quantum eraser experiment}\label{quantumeraser}

We now consider the experiment where each photons path can be "marked" by placing frequency filters {\it before} the beam-splitter, a procedure that will destroy the previously observed beating and, depending on the degree of distinguishability between the photons, will lead to the appearance of dips with different visibility in the Hong, Ou and Mandel experiment \cite{HOM}. There are different ways to implement a quantum eraser experiment with the HOM interferometer, using as a marker the polarization \cite{markedpath,Rempe}.
If the filtered frequencies are such that their difference, $\abs{\omega_1-\omega_2}$ is smaller that the width $\sigma$ of the filters,  this "path-marking" will not be totally effective and the two photons will spatially interfere after the beam-splitter. More specifically, as detailed in the following, the dip's  visibility will depend on the ratio $\abs{\omega_1-\omega_2}/\sigma$, which is proportional to the distance between the distributions representing the ``dead" and ``alive" state of the cat.

\begin{figure*}
\includegraphics[width=1.0\textwidth]{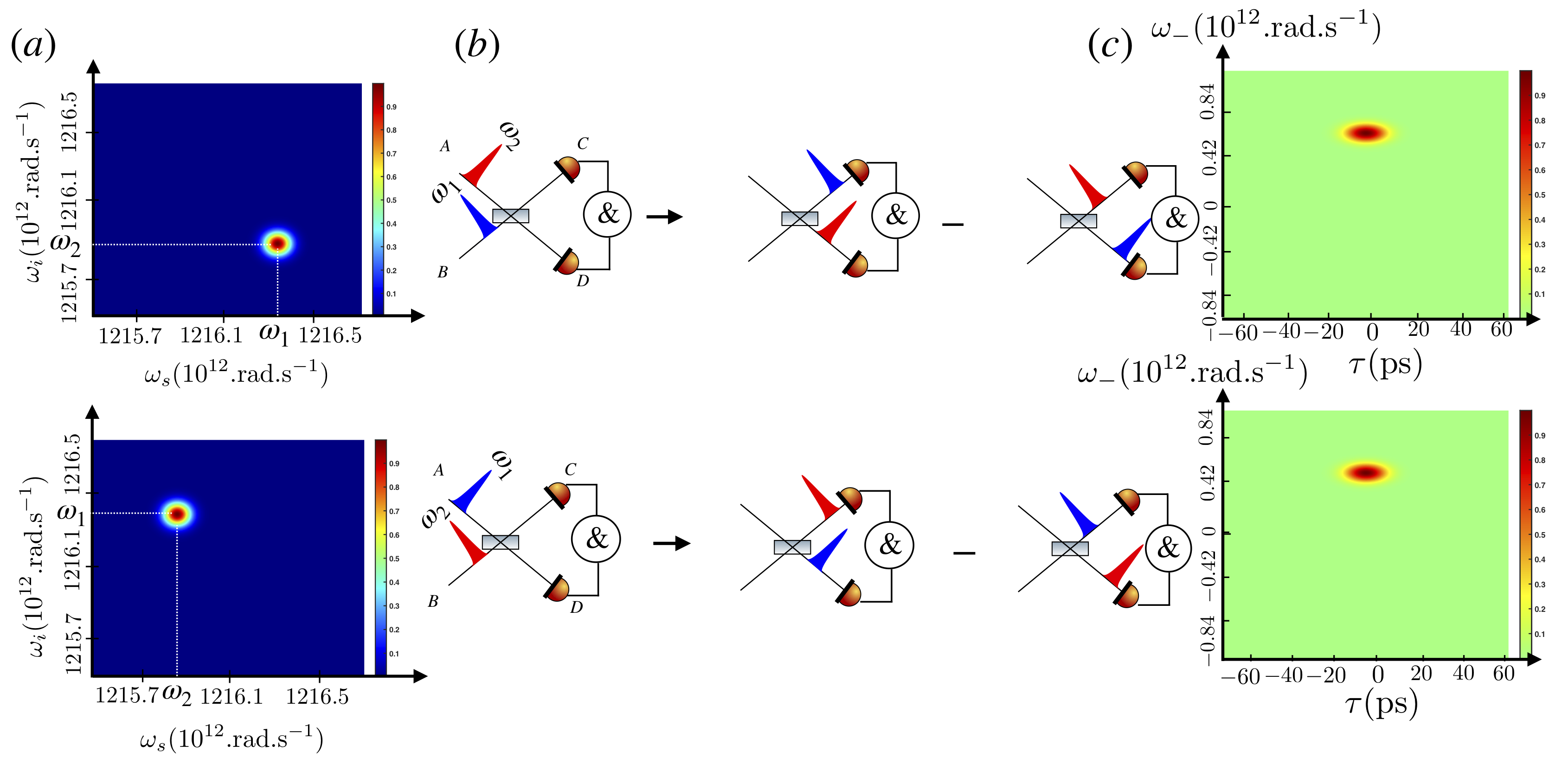}
\caption{\label{diag}(a) Joint Spectral Intensity of bicolor photon when the frequencies filters are placed  before the beam-splitter and their corresponding diagrams. (b) Diagrams of the quantum state after the beam-splitter, taking into account only the coincidence events. (c) Chronocyclic Wigner distribution $W_{-}(\tau,\omega_{-})$ of the two quantum state (b) when the time-resolution of the detectors is lower than the temporal size of the wave packet and $\abs{\omega_{1}-\omega_{2}}\neq 0$. The distribution is zero when $\omega_{1}\gg \omega_{2}$. }
\end{figure*}

Fig.~\ref{diag} (a), (b) reports the results of the numerical simulations of the chronocyclic Wigner distribution for an initial state (Eq.~(\ref{JSA})) with two frequencies filters placed before the beam-splitter. The analytical expression for the coincidence probability, demonstrated in the appendix \ref{quantumeraserappendix}, is given by:
\begin{equation}\label{earcat}
I(\tau)=\frac{1}{2}(1-e^{-(\omega_{1}-\omega_{2})^{2}/(2\sigma^{2})}\cdot e^{-\tau^{2}\sigma^{2}/2}).
\end{equation}
Expression (\ref{earcat}) evidences that the larger the ratio $(\omega_1-\omega_2)/\sigma$ the lower the visibility is, since the action of the filters makes the photons more distinguishable.

The described situation corresponds to selecting one of the two possible states of signal and idler described in the previous section, i.e., the one where $\omega_s=\omega_1$ and  $\omega_i=\omega_2$. Thus, it corresponds to a Gaussian state in phase space, say, where the cat is alive. As shown in \cite{Tom}, the HOM coincidence measurement corresponds to a cut along the frequency axis in phase space. When $\abs{\omega_1-\omega_2} > \sigma$, the two photons have no spectral overlap, explaining a visibility that reaches  $1/2$, which is the classical statistical (uncorrelated) situation. In other words, the more signal and idler become distinguishable, the more the Wigner distribution corresponds to the one of a quantum superposition of two distinguishable states, since the Gaussians corresponding to each one of them become farther from the origin. 
In order to recover the dip representing the Gaussian chronocyclic Wigner distribution of a ``quasi-classical" state,  one possibility is using Electro Optical Modulators (EOM). But from our analogies, something simpler can be conceived. We can think, for instance, of displacing the filters central frequencies such that $\omega_1 \rightarrow \omega_1-\mu$ and $\omega_2$ remains constant. Then, the exponent in the exponential in Eq.~(\ref{earcat}) would be transformed as $\omega_1-\omega_2 \rightarrow \omega_1-\omega_2-\mu$. Varying $\mu$ is a way to displace the frequency axis so that eventually it reach the central point of the two-photon frequency distribution, given by $\mu=\omega_1-\omega_2$. In this case, we recover photon indistinguishability, as if we had translated the origin of the frequency measurement to the central point of the Gaussian representing one of the two possible classical states, as is usually done in experiments. 

To conclude this section, we described in details a method to realize a quantum eraser experiment. It permits to distinguish one of the two possible states that form the Schr\"odinger's cat-like state at the origin of the interference fringes described in Sec.~\ref{section2}. What's more, this method can be generalized to measure the state described by a Gaussian in collective frequency-time phase space centered at the frequency $\mu=\omega_2-\omega_1$. For such, one could simply change the frequency filters of the previous paragraphs, so that $\omega_s=\omega_2$ and $\omega_i=\omega_1$.

\section{Analogy between the time-resolved HOM experiment and a biphoton Young's experiment}\label{sectionthree}

In this section, we will provide an analogy between the time-resolved HOM experiment and the Young two-slits experiment in the case of two anti-correlated photons in position. 

\subsection{Effect of the time resolution of the photodetector}\label{effecttime}
In the previous section, the temporal size of the wavepacket $\delta$ was much smaller than the integration time $T$ of the photodetector. In this section we investigate the other limit $\delta \gg T$  and reinterpret the measured spatial beating in coincidence measurement presented in the Refs \cite{Rempe,Rempe2} in terms of the signature of a frequency-time cat-like state.
In this limit, the probability of the joint detection is studied as a function of two temporal parameters, the optical delay between the two arms $\tau$ and the time-difference $\overline{\tau}$ between two detections: 
\begin{equation}\label{coinresol}
I(\tau,\overline{\tau})=\int \text{d}t_{0} \abs{\bra{t_{0},t_{0}+\overline{\tau}}\ket{\psi_{\tau}}}^{2}. 
\end{equation}
The full derivation of this result is presented in the Appendix \ref{timeresolution}. $t_{0}$ is the time of the first detection and $\ket{\psi_{\tau}}$ is the state after the beam-splitter:
\begin{equation}
\ket{\psi_{\tau}}=\frac{1}{2}\iint \text{d} t_{s} \text{d} t_{i} (\text{JTA}(t_{s}-\tau,t_{i}) - \text{JTA}(t_{i}-\tau,t_{s})) \ket{t_{s},t_{i}},
\end{equation}
where JTA stands for Joint Temporal Amplitude and is the Fourier transform of the JSA. The coincidence detection probability given by Eq.~(\ref{coinresol}) becomes:
{\small{\begin{equation}\label{coincidenceafterbs}
I(\tau,\overline{\tau})=\frac{1}{4}[\int \text{d}t_{0} \abs{\text{JTA}(-\tau+t_{0}, t_{0}+\overline{\tau})-\text{JTA}(-\tau+t_{0}+\overline{\tau},t_{0})}^{2}].
\end{equation}}}
 We point out that for a vanishing time difference $\overline{\tau}$, the coincidence detection for any optical time delay and any wave function is zero, a situation which has no equivalent in the experiments described in the present article and is analyzed in \cite{Rempe2}. \\

\begin{figure}[h!]
\includegraphics[width=0.35\textwidth]{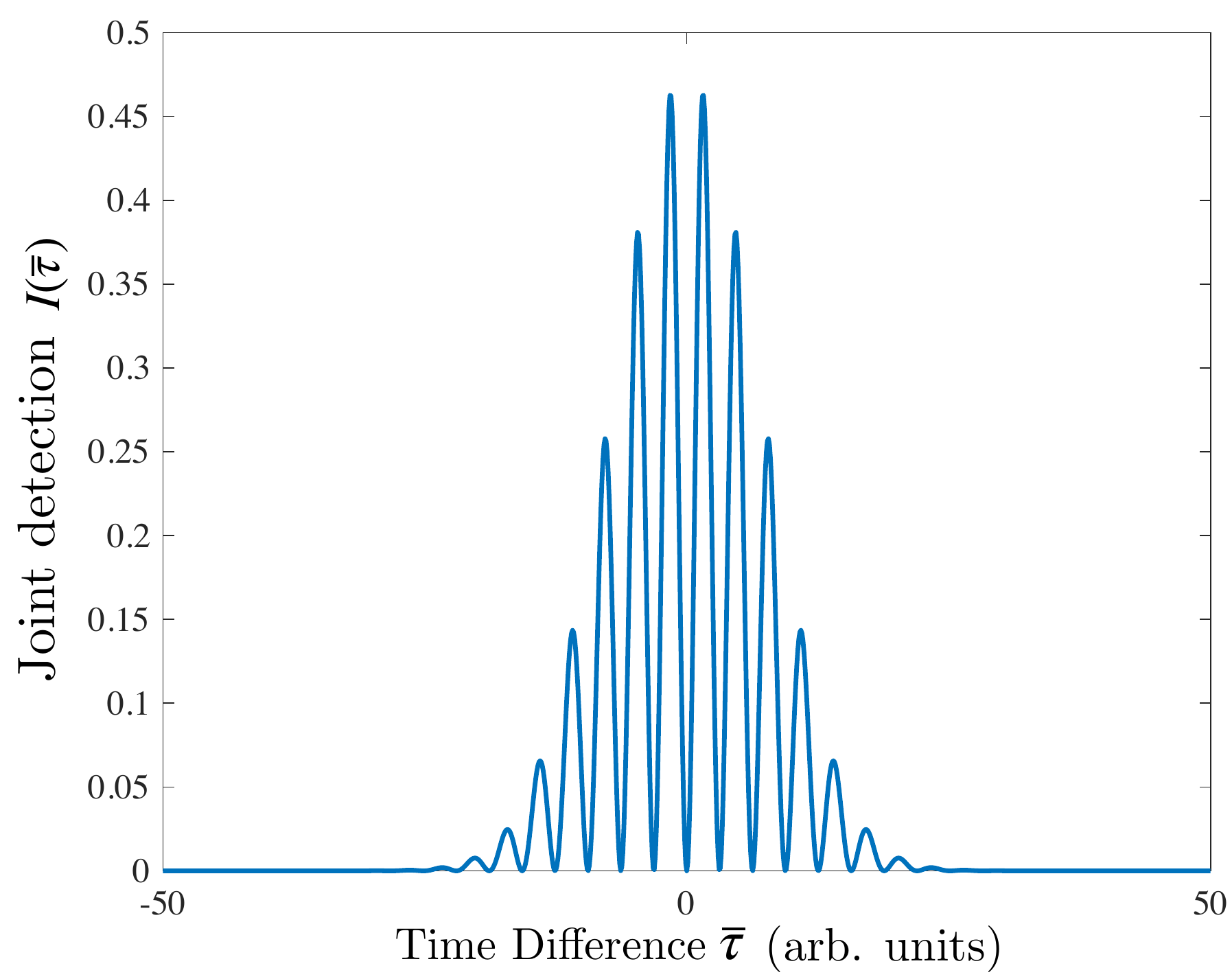}
\caption{\label{catoscillation} Joint detection probability $I(\overline{\tau})$ as a function of the time difference $\overline{\tau}$ in arbitrary units. It corresponds to the marginal of the chronocyclic Wigner distribution of a frequency (odd) cat state and differ from the non-resolved HOM experiment (see Fig.~\ref{fig: chathom1} (d)) where the cut of the chronocyclic Wigner distribution is actually obtained.}
\end{figure}

We now consider again the experiment where two frequency filters are placed before the beam-splitter, as in the quantum eraser experiment described in Sec.~\ref{quantumeraser}. When the optical path delay is set to zero, $\tau=0$, the probability of the joint detection as a function of the time-difference $\overline{\tau}$ detection shows a spatial beating, as shown in Fig.~4 in Ref.~\cite{Rempe2}. 
We now explain the physical reason of this beating and interpret it again as the experimental evidence of a frequency-time cat-like state. We start by giving the analytic expression of the coincidence detection probability. The expression of the JTA for two frequency anti-correlated photons, when two spectral filters are placed before the beam-splitter is:
\begin{equation}\label{JTAcorr}
\text{JTA}(\overline{\tau})=e^{-(\omega_{1}-\omega_{2})^{2}/4\sigma^{2}}e^{-\overline{\tau}^{2}\sigma^{2}/4}e^{-i\overline{\tau}(\omega_{1}-\omega_{2})},
\end{equation}
where the JTA depends only on the time difference  $\overline{\tau}=t_{s}-t_{i}$ owing to the anti-correlation in frequencies. Taking into account Eq.~(\ref{coincidenceafterbs}), the joint detection measurement  $I(\overline{\tau})\propto\abs{\text{JTA}(\overline{\tau})-\text{JTA}(-\overline{\tau})}^{2}$ becomes:
\begin{equation}
I(\overline{\tau})=\frac{1}{4}[\abs{\text{JTA}(\overline{\tau})}^{2}+\abs{\text{JTA}(-\overline{\tau})}^{2}-2\text{Re}(\text{JTA}(\overline{\tau})\text{JTA}^{*}(-\overline{\tau}))].
\end{equation}
Finally using Eq.~(\ref{JTAcorr}) and after normalization, we obtain:
\begin{equation}
I(\overline{\tau})=\frac{1}{2}e^{-\overline{\tau}^{2}\sigma^{2}}(1-\text{cos}(2\overline{\tau}(\omega_{1}-\omega_{2})).
\end{equation}
 Alternatively the joint detection can be written as $I(\overline{\tau})=\abs{\bra{\overline{\tau}}\ket{\psi}}^{2}$, where the probability amplitude $\bra{\overline{\tau}}\ket{\psi}$ is the Fourier transform of the sum of the spectral function of the filters. This measurement gives the marginal of the chronocyclic Wigner distribution, defined in the appendix Eq.~(\ref{Wignerchro}), $\int \text{d}\omega W_{-}(\omega,\overline{\tau})=I(\overline{\tau})$ and in that sense is perfectly analogous to the intensity measured in a biphoton Young experiment as we shall see in the next section, except for the sign in front of the interference term which comes from the reflectivity of the beam-splitter. The joint detection measurement is represented on Fig.~\ref{catoscillation}. It shows an oscillatory behavior, which is the marginal of the chronocyclic Wigner distribution of a frequency-time cat-like state along the time axis.

\subsection{Double-slit experiment with a biphoton state}\label{sectionthreeB}

We now develop the analogy between the biphoton Young's experiment and the time-resolved HOM experiment. This analogy is useful to understand the previous experiment in terms of frequency-time cat-like state. Similar experiment to the one described below can be found in Refs. \cite{biphotonyoung,biphotonyoung1}.

\begin{figure}[h!]
\includegraphics[width=0.49\textwidth]{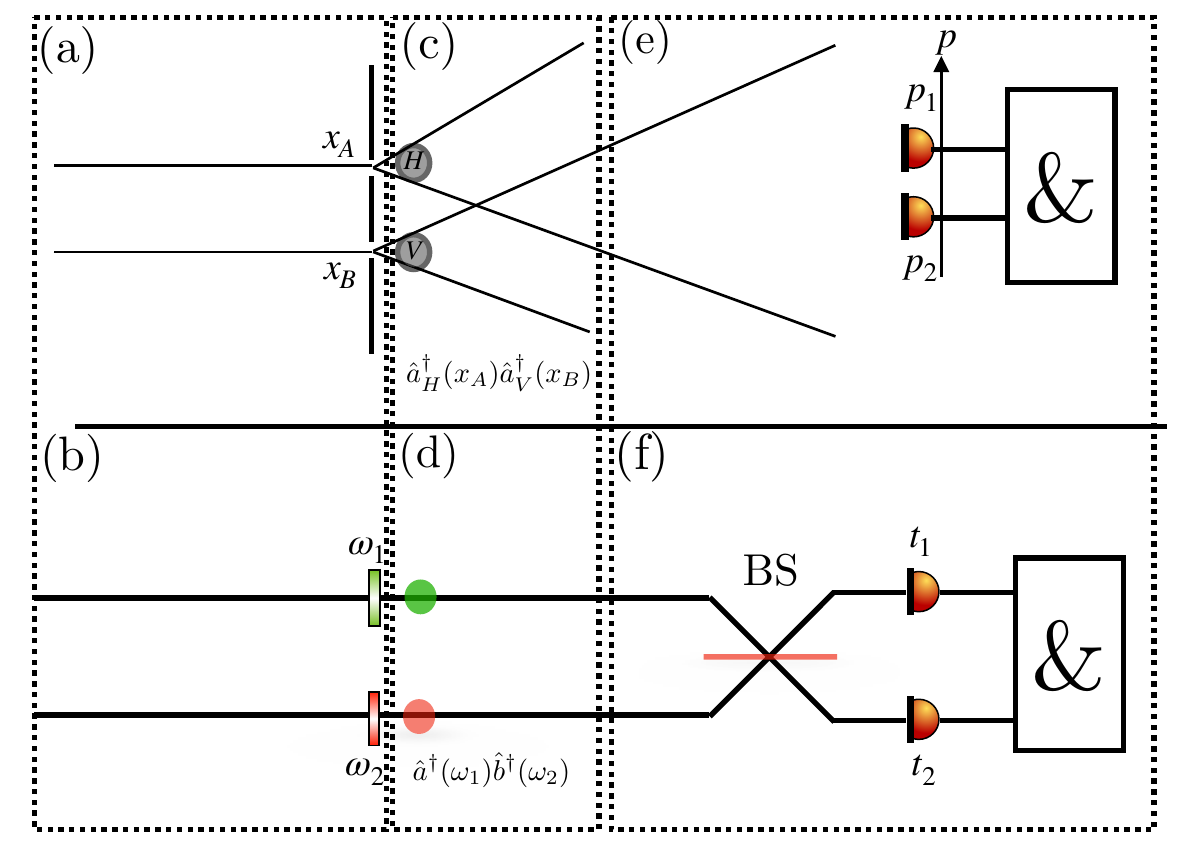}
\caption{\label{Youngexp} Analogy between the biphoton Young's experiment and the time-resolved Hong Ou and Mandel interferometer. (a) A biphoton with polarization H/V crosses the slits centered at $x_{A}/x_{B}$.  (b) A biphoton state in port A/B cross the filter centered at frequency $\omega_{1}/\omega_{2}$.  (c) Propagation of the biphoton in the near field. (d) Propagation of the biphoton  (the dispersion in free space being neglected). (e, f) An even (resp. odd) cat state is produced in the far field (resp. after the beam-splitter (BS in the figure) ) where joint detection measurement is performed. The detection parameters are noted in the two experiments $p_{1/2}=kx_{1/2}/z$ and $t_{1/2}$ but only their difference matter.}
\end{figure}

We consider two position anti-correlated photons which are polarized and sent through two slits: a vertically (V) polarized photon can pass only in the lower slit and a horizontally (H) photon in the upper slit as indicated in Fig.~\ref{Youngexp} (c).  The spatial degree of freedom in the HOM experiment corresponding to the two ports of the interferometer is translated in the Young's experiment into the polarization since they both constitute discrete degrees of freedom. 
In the far field regime using the Fraunhofer approximation, propagation leads to the Fourier transform of the photonic transverse spatial  variables \cite{singlephotonYoung,Huygens}.  Thus, the two slits experiment with anti-correlated photons  leads to the generation of position-momentum cat states without post-selection (see Fig.~\ref{Youngexp}).

 Formally, the two-photon cat state can be described by the wave function after the two slits:
 \begin{multline}
  \ket{\psi}=\iint \text{d}x_{1} \text{d}x_{2} F(x_{1},x_{2})f_{x_{A}+x_{B}}(x_{+}) \\\cross f_{x_{A}-x_{B}}(x_{-})\hat{a}^{\dagger}_{H}(x_{1})\hat{a}^{\dagger}_{V}(x_{2})\ket{0},
\end{multline}
 where $F(x_{1},x_{2})$ is the transverse distribution of the two photon state before the two slits. $\hat{a}^{\dagger}_{H/V}(x_{i})$ is the creation operator of a single photon at position $x_i$ with polarization $H/V$. The two slits behave as a position filter that can be modeled  by a Gaussian function $\overline{f_{x_{A/B}}}(x)=\text{exp}(-(x-x_{A/B})^{2}/(2\sigma^{2}))$, with $\sigma$ the spatial width of the filter which is analogous to the width of the frequency filter as in Eq.~(\ref{filter}). We also employed the notation used for the factorized form of the function filters $\overline{f_{x_{A}}}(x_{1})\overline{f_{x_{B}}}(x_{2})= f_{x_{A}+x_{B}}(x_{+})f_{x_{A}-x_{B}}(x_{-})$ where $x_{\pm}=x_{1}\pm x_{2}$. We will consider again the factorization $F(x_{1}, x_{2})=\delta(x_{+})f_{-}(x_{-})$ and the condition $\sigma \ll \Delta$, where $\Delta$ is the width of the slit of $f_-$. In the near field, we observe two Gaussian peaks which are the slits transmittance.\\  
In the far-field, two detectors are placed at position $x_{i}$, $i=1,2$ and at a distance from the slits $z$. Propagation in the far field  ($z \gg kx_{1}^{2}$) plays the role of the beam-splitter: due to the diffraction, each detector detects photons coming from either slits, forming a coherent superposition of both possible polarized photons that propagated until $z$.
With the same calculation of the previous section, the joint detection $I(\overline{p})=\abs{\bra{\overline{p}}\ket{\psi}}^{2}$ is: 
\begin{equation}
I(\overline{p})=\frac{1}{2}e^{-\overline{p}^{2}\sigma^{2}}(1+\text{cos}(2\overline{p}(x_{A}-x_{B})),
\end{equation}
where $\overline{p}=k(x_{1}-x_{2})/z$ which shows the signature of the creation of an even position-momentum cat. Again the amplitude of probability $\bra{\overline{p}}\ket{\psi}$ corresponds to the coherent sum of the Fourier transform of the transmittance of the slits.
This biphoton Young experiment can be considered as a "momentum resolved" detection scheme, where the detection parameter $\overline{p}$ is analogous to the time difference $\overline{\tau}$ in the time-resolved HOM experiment.  \\
 The propagation can also be viewed as a $\pi/2$ rotation in the position-momentum phase space, or a Fourier transform. Accessing the marginals of the Wigner distribution between these two limits corresponds to implementing fractional Fourier transform to the state.

\section{Manipulating the frequency-time degree of freedom of the biphoton state}\label{sectionfour}

 In this last section, we discuss a method to perform a fractional Fourier transform in frequency-time variable based on an experimental technique that can be implemented in a chip in transverse pump configuration \cite{Orieux2, toolbox}. We will work in the non time-resolved detection limit $\delta \gg T$.

The aim of this section is to provide the analogous phase gate in position-momentum variable $\hat{P}(\theta)=\text{exp}(i\theta \hat{X}^{2})$ (where $\theta$ is the rotation angle and $\hat{X}$ is the position operator) in the collective frequency-time variable $(\omega_{-},\tau)$ of the biphoton state. We propose a solution using the integrated photonic chip described in Sec.~\ref{reinterpretation} and \cite{integrated1} by engineering the phase matching function so as to perform a rotation in the frequency-time phase space.\\
If the length of the circuit is much greater than the spatial width of the pump profile, the phase matching function is the Fourier transform of the spatial profile of the pump $A(z)$ \cite{toolbox,integrated1}:
\begin{equation}\label{fractio}
f_{-}(\omega_{-})=\int_{\mathds{R}} A(z) e^{i\omega_{-}z/v_{g}}\text{d}z,
\end{equation}
where $v_{g}$ is the group velocity and $z$ is the direction of the propagation of the two photons. According to \cite{toolbox} and Eq.~(\ref{fractio}) the spatial profile of the pump is the rescaled Fourier transform of the phase matching function $\tilde{f}(t_-)=A(t_{-}v_{g})$. The adapted spatial profile to perform the rotation in the frequency-time phase space is:
\begin{equation}
A(z)=e^{-\frac{z^{2}}{2\Delta z^{2}}}e^{iz^{2}/a^{2}},
\end{equation}
where $\Delta z$ is the spatial width of the pump. The term $e^{iz^{2}/a^{2}}$ can be obtained using a spatial light modulator (SLM)  to produce a quadratic spatial phase with a curvature $a$ that can be experimentally controlled and varied. It can be viewed as a time chirp owing to the relation Eq.~(\ref{fractio}). This frequency-time gate can also be considered as a time lens \cite{timelens}. \\
The chronocyclic Wigner distribution of the phase matching function is modified as follows: 
\begin{equation}
W_{-}(\omega_{-},\tau)\rightarrow W_{-}(\omega_{-}-\tau a,\tau).
\end{equation}
Since the coincidence probability is the measurement of the cut at $\omega_{-}=0$ of the Wigner distribution, this solution enables the measurement of the full distribution  without a frequency shift but rather by controlling the parameter $a$. Then, changing the value of  $a$ and repeating the OM experiment by varying  $\tau$, we can measure different cuts of the chronocyclic Wigner distribution corresponding to different frequency values.\\
For integrated or bulk optical system, this phase gate can be realized with a grating, which map frequency to spatial degree of freedom followed by a SLM and then another grating to return in the frequency degree of freedom domain \cite{cubicHOM}.\\

\section{Conclusion, perspectives}\label{conclusion}

As a conclusion, we fully revisited the OM experiment in terms of Chronocyclic Wigner distribution in frequency-time phase space for non resolved time coincidence detection. In the new interpretation of the experiment proposed here, interference fringes are associated to a signature of a measurement of a two photon frequency-time Schr\"odinger cat-like state. An analogy between the time-resolved experiment and the biphoton Young's experiment has also been developed, which can help to design experiments.
Lastly, we propose several approaches to engineer different frequency-time cat-like state for application in quantum metrology. Due to the non-commutativity structure of the frequency-time phase space at the single photon level, these variables are equivalent to the quadrature position-momentum continuous variable.  Another perspective of our work is applying it on quantum information protocols based on continuous variables by using the mathematical analogy between single mode/many photon states (field quadratures) and many modes/single photon ones \cite{fabregkp}.  Our results can hence be applied to propose non-locality tests based on the detection of entangled Schr\"odinger cat states \cite{MOI}, on the coding of quantum information in Schr\"odinger cat states \cite{catcode} and other parity-based continuous variables quantum protocols. Our work provides a new perspective on quantum state engineering and measurement, opening applications in quantum information and metrology. For instance, in  \cite{biphotoncolor}, the authors use a frequency-time cat state produced without post-selection in a metrology protocol to measure the thermal dilatation of an optical fiber.

\section*{ACKNOWLEDGMENT}
The authors gratefully acknowledge ANR (Agence Nationale de la Recherche)  for  the  financial  support  of this work through Project SemiQuantRoom (Project No.ANR-14-CE26-0029) and through Labex SEAM (Science and  Engineering  for  Advanced  Materials  and  devices) project  ANR 11 LABX 086, ANR 11 IDEX 05 02. ANR and CGI (Commissariat à l’Investissement d’Avenir) are gratefully acknowledged for their financial support of this work through Labex SEAM (Science and Engineering for Advanced Materials and devices), ANR-10-LABX-0096 and ANR-18-IDEX-0001". The authors acknowledge financial support by ANR/CNPq HIDE and  ANR COMB. J. Belhassen acknowledges support from Delegation Generale de l'Armement.

\appendix

\section{FREQUENCY-TIME CAT-LIKE STATE}

\subsection{Chronocyclic Wigner distribution of frequency-time cat-like state}\label{Wignercatt}

The chronocyclic Wigner distribution of a general two-photon state is defined as \cite{Silberhorn,fabregkp}:
{\small{\begin{multline}
W(\omega_{s},\omega_{i},t_{s},t_{i})=\iint \text{d}\omega' \text{d}\omega'' e^{2i\omega't_{s}}e^{2i\omega''t_{i}}\\\bra{\omega_{s}-\omega',\omega_{i}-\omega''}\hat{\rho}\ket{\omega_{s}+\omega',\omega_{i}+\omega''}.
\end{multline}}}
The construction of this frequency-time distribution was shown in \cite{fabregkp} and emphasizes the quantumness of this distribution when the source of light is composed of single photon. For a pure state $\hat{\rho}=\ket{\psi}\bra{\psi}$, and with a wave function of the form:
\begin{equation}\label{wavefunctionsep}
\ket{\psi}=\iint f_{+}(\omega_{+})f_{-}(\omega_{-}) \ket{\omega_{s},\omega_{i}} \text{d}\omega_{s}\text{d}\omega_{i}.
\end{equation}
we obtain the chronocyclic Wigner distribution:
{\small{\begin{multline}
W(\omega_{s},\omega_{i},t_{s},t_{i})=\iint \text{d}\omega' \text{d}\omega'' e^{2i\omega't_{s}}e^{2i\omega''t_{i}} f_{+}(\omega_{+}-\omega'-\omega'')\\f_{+}^{*}(\omega_{+}+\omega'+\omega'')f_{-}(\omega_{-}-\omega'+\omega'')f_{-}^{*}(\omega_{-}-\omega''+\omega'),
\end{multline}}}
After performing a change of variable, the chronocyclic Wigner distribution can be factorized as follows:
\begin{equation}
W(\omega_{s},\omega_{i},t_{s},t_{i})=W_{+}(\omega_{+},t_{+})W_{-}(\omega_{-},t_{-}),
\end{equation}
where we have noted,
\begin{equation}\label{Wignerchro}
W_{\pm}(\omega,t)=\int \text{d}\omega' e^{2i\omega' t} f_{\pm}(\omega-\omega')f_{\pm}^{*}(\omega+\omega').
\end{equation}
As it was shown in \cite{Tom}, the measured coincidence probability in the generalized HOM experiment is related to the chronocyclic Wigner distribution of the phase matching function:
\begin{equation}\label{coinWigner}
I(2\tau,\omega_{-})=\frac{1}{2}(1-W_{-}(\omega_{-},\tau)).
\end{equation}
The relevant part of the chronocylic Wigner distribution for this experiment is the $W_{-}$ part. For an even (resp. odd) frequency-time cat-like state, with a phase matching function of the form: 
\begin{equation}\label{coherentsuperposition}
f_{-}(\omega_{-})=\frac{1}{\sqrt{2}}(f_{\omega_{1}-\omega_{2}}(\omega_{-})\pm f_{\omega_{2}-\omega_{1}}(\omega_{-})),
\end{equation}
where $f_{\omega_{i}}(\omega_{-})=\text{exp}(-(\omega_{-}-\omega_{i})^{2}/(2\sigma^{2}))$, the chronocylic Wigner distribution is:
{\small{\begin{multline}\label{Wignercat}
W_{-}(\omega_{-},\tau)=\frac{1}{2}[\int \text{d}\omega' e^{2i\omega' \tau} (f_{\omega_{1}-\omega_{2}}(\omega_{-}+\omega')+f_{\omega_{2}-\omega_{1}}(\omega_{-}+\omega'))\\\cross(f^{*}_{\omega_{1}-\omega_{2}}(-\omega'+\omega_{-})+f^{*}_{\omega_{2}-\omega_{1}}(-\omega'+\omega_{-}))].
\end{multline}}}
The distribution is composed of three terms, 
{\small{\begin{multline}\label{Wignercat}
W_{-}(\omega_{-},\tau)=W_{\text{cat}}(\omega_{-},\tau)=W_{12}(\omega_{-},\tau)+W_{21}(\omega_{-},\tau)\\+ W_{\text{beating}}(\omega_{-},\tau).
\end{multline}}}
The shape of the Wigner distribution of the  frequency-time cat-like state $W_{\text{cat}}(\omega_{-},\tau)$ is showed in the main text on Fig.~\ref{fig: chathom1}. The first and second terms are Gaussian functions centered at $\omega_{-}=\omega_{1}\mp \omega_{2}$ and $\tau=0$:

{\small{\begin{multline}\label{catf2}
W_{12}(\omega_{-},\tau)=\int d\omega' e^{2i\omega' \tau}f_{\omega_{1}-\omega_{2}}(\omega_{-}+\omega')f^{*}_{\omega_{1}-\omega_{2}}(-\omega'+\omega_{-})
\\=e^{-\tau^{2}\sigma^{2}}e^{-(\omega_{-}-(\omega_1+\omega_{2}))^{2}/\sigma^{2}}.
\end{multline}}}
 Whereas the interference term is under the form:
{\small{\begin{multline}\label{Wignercatbeat}
 W_{{\small{\text{beating}}}}(\omega_{-},\tau)=\int e^{2i\omega' \tau}(f_{\omega_{1}-\omega_{2}}(\omega_{-}+\omega')f^{*}_{\omega_{2}-\omega_{1}}(-\omega'+\omega_{-})\\+f_{\omega_{2}-\omega_{1}}(\omega_{-}+\omega')f_{\omega_{1}-\omega_{2}}^{*}(-\omega'+\omega_{-}))\text{d}\omega' .
\end{multline}}}
Since the filters are modeled by an even function, we can write the product as: $f_{\omega_{1}-\omega_{2}}(\omega_{-}+\omega')f^{*}_{\omega_{2}-\omega_{1}}(-\omega'+\omega_{-})=f_{\omega_{1}-\omega_{2}}(\omega_{-}+\omega')f^{*}_{\omega_{1}-\omega_{2}}(\omega'-\omega_{-})$. Without a post-selection procedure, the coincidence measurement with a HOM interferometer permits to access to the cut of the beating term:
\begin{equation}
 W_{\text{beating}}(0,\tau)= e^{-\tau^{2}\sigma^{2}}\text{cos}(2\tau(\omega_{2}-\omega_{1})).
\end{equation}
The terms  $W_{12}$ and $W_{21}$ can not be access through this measurement because they are not along the $\omega_{-}=0$ axis, but could be obtained thanks to the generalized HOM experiment \cite{Tom}.

\section{Quantum erasing experiment}\label{quantumeraserappendix}
Here we detail the coincidence probability measured with the HOM interferometer when the frequency filters are placed before the beam-splitter (see Sec.~\ref{quantumeraser}). We start from the state defined by Eq.~(\ref{wavefunctionsep}). After the filtering operation the state becomes:

\begin{equation}
\ket{\psi}=\iint f_{+}(\omega_{+})f_{-}(\omega_{-}) \overline{f_{\omega_{1}}}(\omega_{s})\overline{f_{\omega_{2}}}(\omega_{i}) \ket{\omega_{s},\omega_{i}} \text{d}\omega_{s} \text{d}\omega_{i}.
\end{equation}
Using the relation $\overline{f_{\omega_{1}}}(\omega_{s})\overline{f_{\omega_{2}}}(\omega_{i})=f_{\omega_{1}-\omega_{2}}(\omega_{-})f_{\omega_{1}+\omega_{2}}(\omega_{+})$, the condition $f_{+}(\omega_{+})=\delta(\omega_{+})$ and considering again than the frequency width of the phase matching function is larger than the width of the filters, the wave function $\ket{\psi_{\tau}}$ can be written before the beam-splitter as:
\begin{equation}
\ket{\psi_{\tau}}=\int \text{d}\omega_{-} e^{i\omega_{-}\tau}f_{\omega_{1}-\omega_{2}}(\omega_{-}) \ket{\omega_{-},-\omega_{-}},
\end{equation}
and we considered the following normalization: $\int \text{d}\omega_{-} \abs{f_{\omega_{1}-\omega_{2}}(\omega_{-})}^{2}=1$.  After the beam-splitter and post-select only the coincidence terms, the wave function becomes
\begin{equation}
\ket{\psi_{\tau}}=\frac{1}{2} \int \text{d}\omega_{-} e^{i\omega_{-}\tau}f_{\omega_{1}-\omega_{2}}(\omega_{-}) (\ket{\omega_{-},-\omega_{-}}-\ket{-\omega_{-},\omega_{-}}).
\end{equation}
We then performe a change of variable:
{\small{\begin{multline}
\ket{\psi_{\tau}}=\frac{1}{2} \int \text{d}\omega_{-} (e^{i\omega_{-}\tau}f_{\omega_{1}-\omega_{2}}(\omega_{-})-e^{-i\omega_{-}\tau}f_{\omega_{1}-\omega_{2}}(-\omega_{-}))\\\cross \ket{\omega_{-},-\omega_{-}}.
\end{multline}}}
 Once  the coincidence  is measured $I(\tau)=\iint \text{d}\omega_{s} \text{d}\omega_{i} \bra{\omega_{s},\omega_{i}}\hat{\rho}_{\tau}\ket{\omega_{s},\omega_{i}}$, with $\hat{\rho}_{\tau}=\ket{\psi_{\tau}}\bra{\psi_{\tau}}$, we point out that the first two terms do not depend on $\tau$ and the two crossed terms do not give rise to a beating pattern. Indeed, we obtain:
\begin{equation}
I(\tau)=\frac{1}{2}[1-\text{Re}(\int \text{d}\omega_{-} f_{\omega_{1}-\omega_{2}}(\omega_{-})f_{\omega_{1}-\omega_{2}}^{*}(-\omega_{-})e^{2i\omega_{-}\tau})].
\end{equation}
From this last equation, after performing the integration over $\omega_{-}$, we get Eq.~(\ref{earcat}). We recognize the cut of the Chronocyclic Wigner distribution of $W_{12}$ (or $W_{21}$) at the frequency $\omega_{-}=0$, see Eq.~(\ref{catf2}).

\section{Influence of time resolution of the photo-detector in the HOM experiment}\label{timeresolution}
In this section, we remind the results on the influence of the time resolution of the photodetector from \cite{Rempe,Rempe2}.

\subsection{Probability of the joint detection time detection in the HOM experiment}

The probability to measure one photon at the port C at time $t_{1}$ and one photon at the port D at time $t_{2}$ (see Fig.~\ref{FigureHom} for instance) is given by the second order correlation function: 
\begin{equation}\label{secondorder}
G^{2}(t_{1},t_{2},\tau))=\text{Tr}(\hat{\rho}(\tau)\hat{c}^{\dagger}(t_{1})\hat{d}^{\dagger}(t_{2})\hat{d}(t_{2})\hat{c}(t_{1})),
\end{equation}
 where $\tau$ is again the photon delay.  $\hat{c}(t_{1})$ denotes the creation photon operator at time $t_{1}$ in the spatial port C. Any photodetectors have a finite resolution $T$, then  we access experimentally to  the probability to detect one photon in the interval $t_{0}\pm T$ and the second in the interval $t_{0}'\pm T$:
\begin{equation}
I(t_{0},t'_{0}, \tau)=\int_{t_{0}-T}^{t_{0}+T} \text{d}t_{1}\int_{t'_{0}-T}^{t'_{0}+T} \text{d}t_{2} G^{2}(t_{1},t_{2},\tau).
\end{equation}
When $T\gg \delta$, the range of the integral can be extended to the infinity and we obtain:
\begin{equation}
I(\tau)=\iint \text{d}t_{1}\text{d}t_{2} G^{2}(t_{1},t_{2},\tau).
\end{equation}
 We can rewrite the coincidence probability as :
\begin{multline}
I(\tau)=\iint \text{d}t_{1}\text{d}t_{2} \iint \text{d}t_{3}\text{d}t_{4} \bra{t_{3},t_{4}}\hat{\rho}_{\tau}\hat{c}^{\dagger}(t_{1})\hat{d}^{\dagger}(t_{2})\\
\cross \hat{d}(t_{2})\hat{c}(t_{1})\ket{t_{3},t_{4}}.
\end{multline}
By applying the bosonic operators and after integration, the coincidence detection is $I(\tau)=\iint \text{d}t_{1}\text{d}t_{2} \bra{t_{1},t_{2}}\hat{\rho}_{\tau}\ket{t_{1},t_{2}} $ which can be written alternatively, for a pure state, using the closure relation: $I(\tau)=\iint \text{d}\omega_{s}\text{d}\omega_{i} \abs{\bra{\psi_{\tau}}\ket{\omega_{s},\omega_{i}}}^{2}$, showing the Eq.~(\ref{coincidencegeneral}) in the main text.

We demonstrate now the coincidence probability in the case where $\delta \gg T$:
\begin{equation}
I(\tau,\overline{\tau})=\int G^{2}(t_{0},t_{0}+\overline{\tau},\tau)\text{d}t_{0},
\end{equation}
where $t_{0}$ is the time of the first detection which is integrated, $\tau$  being the optical delay between the two arms and the time-difference between two detections  is noted $\overline{\tau}$. 
Thanks to this last equation, we easily prove Eq.~(\ref{coinresol}).

\subsection{Definition of the probability of the  joint detection in the Young experiment}

We finish this section by reminding the probability of joint detection of one photon of impulsion $p_{1}$ and the second of impulsion $p_{2}$ in the biphoton Young experiment. It is given by the second order correlation function:
\begin{equation}
G^{2}(p_{1},p_{2})=\sum_{\alpha,\alpha'} \text{Tr}(\hat{\rho}\hat{a}^{\dagger}_{\alpha}(p_{1})\hat{a}^{\dagger}_{\alpha'}(p_{2})\hat{a}_{\alpha}(p_{1})\hat{a}_{\alpha'}(p_{2})),
\end{equation}
where $\hat{a}_{\alpha}(p_{i})/\hat{a}^{\dagger}_{\alpha}(p_{i}) $ denotes the bosonic annihilation (resp. creation) operator of a single photon with polarization $\alpha$ and impulsion $p_{i}=kx_{i}/z$. For two photons anti-correlated in position, only the difference of the impulsion $\overline{p}$ is actually relevant. This mathematical analogy permits to conclude about the equivalence of the two experiments.

\bibliography{bibliocat2}

\end{document}